\documentstyle[pra,aps]{revtex}

\begin{document}
\draft
\title{Optimal squeezing, pure states, and amplification of squeezing in resonance
fluorescence}
\author{Peng Zhou\thanks{
Email: peng@qo1.am.qub.ac.uk} and S. Swain\thanks{
Email: s.swain@qub.ac.uk}}
\address{Department of Applied Mathematics and Theoretical Physics, \\
The Queen's University of Belfast, Belfast BT7 1NN, United Kingdom.}
\date{}
\maketitle

\begin{abstract}
It is shown that $100\%$ squeezed output can be produced in the resonance
fluorescence from a coherently driven two-level atom interacting with a
squeezed vacuum. This is only possible for $N=1/8$ squeezed input, and is
associated with a pure atomic state, {\it i.e.}, a completely polarized
state. The quadrature for which optimal squeezing occurs depends on the
squeezing phase $\Phi ,$ the Rabi frequency $\Omega ,$ and the atomic
detuning $\Delta $. Pure states are described for arbitrary $\Phi ,$ not
just $\Phi =0$ or $\pi $ as in previous work. For small values of $N,$ there
may be a greater degree of squeezing in the output field than the input ---
i.e., we have squeezing amplification.
\end{abstract}


\section{Introduction}

Both theoretical and experimental studies have shown that the resonance
fluorescence of a driven atom can serve as a source of nonclassical light.
For example, Carmichael and Walls, and Kimble and Mandel \cite{antib}
predicted that the resonance fluorescence from a single two-level atom
driven by a coherent laser field of low intensity would exhibit photon
antibunching. The prediction has been confirmed in many laboratories %
\onlinecite{antibexp1,antibexp2,antibexp3}. The sub-Poissonian statistics of
the fluorescent photons emitted in a short time interval by a single atom
was also investigated experimentally \cite{mandel}.

There have been many theoretical investigations of squeezing in resonance
fluorescence, both in terms of the total variances and in terms of the
fluctuation spectra of the phase quadratures. Walls and Zoller, and Loudon 
\cite{WP} showed that the total quantum fluctuations in the phase
quadratures of the resonance fluorescence of a driven two-level atom can be
squeezed below the shot-noise limit. Single-mode \cite{RFsq1}, or
frequency-tunable two-mode \cite{PZ} squeezing with a finite bandwidth may
be obtained, depending on the Rabi frequency and detuning. This internally
produced squeezing results in line narrowing in the resonance fluorescence
spectra \onlinecite{RiceCar,RFsq2}. Phase-quadrature squeezing has also been
studied in the presence of an applied squeezed vacuum %
\onlinecite{RFsq2,RFsq3,RFsq4}.

Experimental observation of squeezing in the fluorescence field has proved a
great challenge, one problem being that atomic motion produces phase shifts
which destroy squeezing \cite{detsq}. This difficulty was surmounted in the
recent experimental advances in homodyne detection schemes of the
fluorescent radiation of a single trapped ion reported by Hoffges {\it et al.%
} \cite{antibexp2}. Within the last couple of years, experiments carried out
by Zhao {\it et al.} \cite{phasesp} have found some evidence of squeezing by
measuring the phase-dependent fluorescence spectra of a coherently driven
two-level atom with a long lifetime, stimulating the further exploration of
squeezing in resonance fluorescence. Very recently, squeezing in the
quadrature with phase $\pi /4$ relative to the driving laser was observed
for the first time in the resonance fluorescence of a single two-level atom 
\cite{lu-ba-th98}.

Also very recently, we have found that squeezing in resonance fluorescence
can be greatly enhanced in a frequency-tunable cavity \cite{cavsq}, or in a
squeezed vacuum \onlinecite{RFsq3,RFsq4}. The latter works mainly in the
regime over which anomalous spectra such as hole-burning and dispersive
profiles \cite{ARF} occur, {\it i.e.}, $\Delta =0$ and $\Phi =0$, where
squeezing occurs in the {\it out-phase} quadrature of the fluorescent field.
In this paper we extend the study to the general case, and show that large
squeezing occurs in {\it different phase} quadratures of the fluorescent
field, depending upon the values of the parameters. The large squeezing is
associated with an atomic pure state (a completely polarized state), and
thereby with a large atomic coherence. Perfect fluorescent squeezing may
only take place for the particular squeezing number $N=1/8$.

There is previous evidence that the value $N=1/8$ is special. It has been
shown that large squeezing in resonance fluorescence was produced for this
input squeezed field \cite{RFsq4}, and that also for this value of $N,$ the
sidebands in the resonance fluorescence spectrum have the same linewidth as
in the $N=0$ case \cite{qu-ko-la89}.

\section{Pure state}

Our model consists of a two-level atom with ground and excited states $%
|0\rangle $ and $|1\rangle $, driven by a monochromatic laser field and
damped by a broadband squeezed vacuum. In the frame rotating with the laser
frequency $\omega _{L},$ the Hamiltonian (in units of $\hbar $), is

\begin{equation}
H=\frac{\Delta }{2}\sigma _{z}+\frac{\Omega }{2}\left( e^{i\phi _{L}}\sigma
_{+}+e^{-i\phi _{L}}\sigma _{-}\right) ,  \label{ham}
\end{equation}
where $\sigma _{z}=(|1\rangle \langle 1|-|0\rangle \langle 0|)$ is the
atomic inversion operator, $\sigma _{+}=|1\rangle \langle 0|$ and $\sigma
_{-}=|0\rangle \langle 1|$ are the atomic raising and lowering operators
respectively, $\Delta =\omega _{A}-\omega _{L}$ is the detuning, $\Omega $
is the Rabi frequency and $\phi _{L}$ is the laser phase. The squeezed
vacuum is characterized by the squeezing photon number $N$, the squeezing
phase $\phi _{s}$, and the strength of the two-photon correlations $M$,
which obeys

\begin{equation}
M=\eta \sqrt{N(N+1)},\;\;\;\;\;\;\;(0\leq \eta \leq 1).
\end{equation}
The value $\eta =1$ indicates an ideal squeezed vacuum, whilst $\eta =0$
corresponds to no squeezing at all --- a thermal field. The squeezed vacuum
may be `turned off' by setting $N=0$. In the remainder of the paper, we take 
$\eta =1,$ and thus $M=\sqrt{N(N+1)}$.

The optical Bloch equations, modified by the squeezed vacuum, are of the form

\begin{eqnarray}
\langle \dot{\sigma}_{x}\rangle &=&-\gamma _{x}\langle \sigma _{x}\rangle
-(\Delta +\gamma M\sin \Phi )\langle \sigma _{y}\rangle ,  \nonumber \\
\langle \dot{\sigma}_{y}\rangle &=&-\gamma _{y}\langle \sigma _{y}\rangle
+(\Delta -\gamma M\sin \Phi )\langle \sigma _{x}\rangle -\Omega \langle
\sigma _{z}\rangle ,  \nonumber \\
\langle \dot{\sigma}_{z}\rangle &=&-\gamma _{z}\langle \sigma _{z}\rangle
+\Omega \langle \sigma _{y}\rangle -\gamma ,  \label{bloch}
\end{eqnarray}
with

\begin{eqnarray}
&&\gamma _{x}=\Gamma +\gamma M\cos \Phi ,  \nonumber \\
&&\gamma _{y}=\Gamma -\gamma M\cos \Phi ,  \nonumber \\
&&\gamma _{z}=\gamma _{x}+\gamma _{y}=2\Gamma ,
\end{eqnarray}
where $\sigma _{x}=(\sigma _{-}+\sigma _{+})$ and $\sigma _{y}=i(\sigma
_{-}-\sigma _{+})$ are the in--phase (X) and out--phase (Y) quadrature
components of the atomic polarization, respectively. (We have assumed in
obtaining eq. (\ref{bloch}) that the applied EM field is coupled to the $%
\sigma _{x}$ component of the atom.) $\Phi =2\phi _{L}-\phi _{s}$ is the
relative phase between the laser field and squeezed vacuum, and $\Gamma
=\gamma \left( N+1/2\right) $, with $\gamma $ the spontaneous decay rate of
the atom into the standard vacuum modes. The modified decay rates of the X
and Y components of the atomic polarization are $\gamma _{x},\,\gamma _{y}$
respectively, whilst $\gamma _{z}$ is the decay rate of the atomic
population inversion.

It has been shown that such a coherently driven two-level atom interacting
with the squeezed vacuum reservoir can collapse into a steady-state which is
a pure state, for the case $\Phi =0$ or $\pi $ \onlinecite{ARF,badc}. This
property is associated with anomalous spectral features in the resonance
fluorescence and probe absorption. The condition for the atom to be in a
pure state is that the quantity $\Sigma ,$ where 
\begin{equation}
\Sigma =\langle \sigma _{x}\rangle ^{2}+\langle \sigma _{y}\rangle
^{2}+\langle \sigma _{z}\rangle ^{2},
\end{equation}
takes the value one \cite{tucc}. We point out here that a steady pure state
can, in fact, be achieved for other values of the squeezing phase as well,
the requirement being that given $\Phi ,$ $\Omega $ and $\Delta $ are chosen
to satisfy $\Sigma =1$. The general pure state has the form

\begin{equation}
|\Psi \rangle =\frac{\sqrt{M}|0\rangle -e^{i\alpha }\sqrt{N}|1\rangle }{%
\left( M+N\right) ^{1/2}}.  \label{pur}
\end{equation}
where 
\begin{equation}
\alpha =\arctan \left( \frac{\Gamma +\gamma M\cos \Phi }{\Delta +\gamma
M\sin \Phi }\right) .  \label{opt}
\end{equation}
The conditions for the pure state (\ref{pur}) for a few specific cases are
given below:

\begin{equation}
\Phi =0,\hspace{0.5cm}\Delta =0,\hspace{0.5cm}\Omega =\frac{\gamma \sqrt{M}}{%
\sqrt{N+1}+\sqrt{N}},  \label{cond1}
\end{equation}
\begin{equation}
\Phi =\frac{\pi }{2},\hspace{0.5cm}\Delta =\Gamma -\gamma M,\hspace{0.5cm}%
\Omega =\frac{\gamma \sqrt{2M}}{\sqrt{N+1}+\sqrt{N}},  \label{cond2}
\end{equation}
\begin{equation}
\Phi =\pi ,\hspace{0.5cm}\Delta \gg \Gamma -\gamma M,\hspace{0.5cm}\Omega =%
\frac{2\Delta \sqrt{M}}{\sqrt{N+1}-\sqrt{N}}.  \label{cond3}
\end{equation}
Notice that for resonant excitation, a pure state is only possible if $\Phi
=0.$ In general, the pure state (\ref{pur}) describes a completely polarized
atom with the Bloch vector ${\bf B}$ lying on the Bloch sphere with polar
angles $\alpha $ and $\beta $,

\begin{equation}
{\bf B}=\cos \alpha \,\sin \beta \,{\bf e}_{x}+\sin \alpha \,\sin \beta \,%
{\bf e}_{y}+\cos \beta \,{\bf e}_{z}
\end{equation}
where 
\begin{equation}
\beta =\arccos \left( -\frac{M-N}{M+N}\right) .  \label{opta}
\end{equation}
See Figure 1. When $\Phi ,\,\Omega $ and $\Delta $ satisfy the condition (%
\ref{cond1}), then $\alpha =\pi /2$, and the atomic Bloch vector
(polarization) is in the Y--Z plane, whereas if the condition (\ref{cond3})
holds, we have $\alpha =0$ and the atom polarizes in the X--Z plane.

\section{Optimal and Maximal Squeezing}

The measurement of the quadrature squeezing spectrum requires the
fluorescent radiation field to be first frequency filtered and then
homodyned with a strong local oscillator field \cite{phasesp}. The squeezing
may also be detected in terms of the total normally-ordered variances of the
phase quadratures in an alternative experimental scheme, where the total
radiation field and the local oscillator are directly homodyned, without
first frequency filtering \onlinecite{mandel,detsq}. In this paper we
interested in the latter quantity, which can be expressed in terms of the
steady state solution of the Bloch equations (\ref{bloch}) as %
\onlinecite{WP,RFsq1}

\begin{equation}
S_{\theta }=\left\langle :\left( \Delta E_{\theta }\right)
^{2}:\right\rangle =1+\langle \sigma _{z}\rangle -\left( \langle \sigma
_{x}\rangle \cos \theta -\langle \sigma _{y}\rangle \sin \theta \right) ^{2},
\label{sstsq}
\end{equation}
where $E_{\theta }=e^{-i\theta }{\cal E}^{(+)}+e^{i\theta }{\cal E}^{(-)}$
is the $\theta $-phase quadrature of the atomic fluorescence field, measured
by homodyning with a local oscillator having a controllable phase $\theta $
relative to the driving laser. $E_{\theta =0}$ and $E_{\theta =\pi /2}$ are
usually the in-phase (X) and out-of-phase (Y) quadratures of the fluorescent
field, respectively. $S_{\theta }$ is the total normally-ordered variance of
the $\theta $-phase quadrature of the fluorescent field. The field is said
to be squeezed when $S_{\theta }<0$. The normalization we have chosen is
such that maximum squeezing corresponds to $S_{\theta }=-0.25$. Eq. (\ref
{sstsq}) implies that the squeezing occurs at large values of the atomic
coherences, $\langle \sigma _{x(y)}\rangle $.

It is not difficult to show that the total normally-ordered variances in the
phase quadrature component of the fluorescent field reach their minimal value

\begin{equation}
S_{\theta _{o}}=1+\langle \sigma _{z}\rangle -\langle \sigma _{x}\rangle
^{2}-\langle \sigma _{y}\rangle ^{2}=\langle \sigma _{z}\rangle (1+\langle
\sigma _{z}\rangle )+1-\Sigma ,  \label{Sth}
\end{equation}
when the quadrature phase $\theta =\theta _{o}$, where

\begin{equation}
\theta _{o}=\arctan \left( \frac{\Gamma +\gamma M\cos \Phi }{\Delta +\gamma
M\sin \Phi }\right) .  \label{opti}
\end{equation}
(Note that only when $S_{\theta _{o}}<0$ is the resonance fluorescence a
noise-squeezed field.) Furthermore, if the atom is in a pure state, {\em i.e.%
}, $\Sigma =1$, then $S_{\theta _{o}}$ reduces to

\begin{equation}
S_{\theta _{o}}^{PS}=\langle \sigma _{z}\rangle (1+\langle \sigma
_{z}\rangle )\leq 0,  \label{pusq}
\end{equation}
showing that maximum squeezing occurs when $\langle \sigma _{z}\rangle
=-1/2. $ Therefore, a completely polarized atom always radiates a
fluorescent field with $\theta _{o}$-phase quadrature squeezing. The
quadrature phase $\theta _{o}$ is same as the longitudinal angle $\alpha $
of the polarized atom in the Bloch sphere.

It is clear from eq. (\ref{Sth}) that the squeezing is maximal, $S_{\theta
_{o}}=-0.25,$ if the equations $\Sigma =1$ and $\langle \sigma _{z}\rangle
=-1/2$ are {\em simultaneously} satisfied. We find that this is possible 
{\em only} for $N=1/8$, and then analytic solution can be found. The
appropriate values of $\Omega $ and $\Delta $ are 
\begin{equation}
\Delta =\frac{t}{4},\qquad \Omega =\frac{1}{4}\sqrt{3\left( 1+t^{2}\right) }%
,\qquad \alpha =\theta _{0}=\arctan \left( \frac{1}{t}\right) 
\end{equation}
where $t=\tan \left( \Phi /2\right) .$ These expressions are consistent with
the values presented in eqs. (\ref{cond1}) -- (\ref{cond3}). It is clear
that $\Delta \rightarrow \infty $ as $\Phi \rightarrow \pi ,.$and then $%
\Omega \rightarrow \sqrt{3}\Delta .$

We comment briefly why maximal squeezing is only possible for a small value
of $N.$ Whilst, for arbitrary values of $N$ and $\Phi $ it is always
possible to find values of $\Omega $ and $\Delta $ which produce a pure
state $\left( \text{satisfy }\Sigma =1\right) $, the condition $\langle
\sigma _{z}\rangle =-1/2$ is much more stringent. This value of $\langle
\sigma _{z}\rangle $ implies that the external fields must not be so strong
as to saturate the atom. This may be seen by examining the expression for $%
\langle \sigma _{z}\rangle +1/2.$ We require 
\begin{equation}
\langle \sigma _{z}\rangle +1/2\equiv \frac{1}{2}\frac{\left( N+1/2+M\cos
\Phi \right) \Omega ^{2}+\left( 2N-1\right) \left( \Delta ^{2}+1/4\right) }{%
\left( N+1/2+M\cos \Phi \right) \Omega ^{2}+\left( \Delta ^{2}+1/4\right)
\left( 2N+1\right) }=0.  \label{sz}
\end{equation}
Clearly, $\langle \sigma _{z}\rangle +1/2>0$ if $N>1/2, $even if $\Omega =0.$
The condition $\Sigma =1$ requires a nonzero value of $\Omega , $and thus if
the conditions $\Sigma =1$ and $\langle \sigma _{z}\rangle +1/2$ can be
satisfied, and it is not obvious from the outset that they can, we must have 
$0<N<\frac{1}{2}$

This argument does not explain why solutions are only possible for the {\em %
particular} value $N=1/8$, but as we have remarked earlier, the value $N=1/8$
appears to be a special one from several points of view \cite
{RFsq4,qu-ko-la89}. For this value of $N,$the two-photon correlation
strength, $M\equiv \sqrt{N\left( N+1\right) }$ is rational: $M=3/8,$and the
principal decay rates are respectively twice and one-half the decay rate in
the absence of the squeezed vacuum: $\gamma _{x}=\gamma ,\gamma _{y}=\gamma
/4,$when $\Phi =0.$

For small values of $N,$ large squeezing in the fluoresence is possible,
even if it is not maximal. We may conclude from eq. (\ref{opti}) that when $%
\Phi =0$ and $\Delta =0$, optimal squeezing in the fluorescent field always
occurs in the out-of-phase (Y) quadrature component, {\em i.e.}, $\theta
_{o}=\pi /2$ \onlinecite{RFsq1,RiceCar,RFsq2,RFsq3,RFsq4}. When $\Phi =\pi /2
$ and $\Delta =\Gamma -\gamma M$, then $\theta _{o}=\pi /4$, and optimal
squeezing takes place in the $\pi /4$-phase quadrature \cite{lu-ba-th98}.
When $\Phi =\pi $ and $\Delta \gg \Gamma -\gamma M$, then $\theta _{o}=0$,
and optimal squeezing is always in the in-phase (X) quadrature %
\onlinecite{PZ,cavsq}.

\subsection{In-phase quadrature squeezing}

We here present a detailed study of the fluorescence squeezing in the case
of $\Phi =\pi$ and $\Delta \neq 0$, which has previously received little
attention. As we know from the above discussion, the squeezing is exhibited
in the in-phase (X) quadrature component.

Figure 2 shows $S_{\theta =0}=S_{X}$, the in-phase quadrature of the
fluorescent field, in a 3D plot against the Rabi frequency $\Omega $ and the
squeezed phase $\Phi $, for $N=0.1,\,\Delta =10\gamma T$ and \thinspace $%
\gamma =1$. Clearly, the greater squeezing occurs for large phases. When $%
\Omega \simeq 16\gamma $ and $\Phi =\pi $, then $S_{X}\simeq -0.25$
displaying the optimal ($100\%$) degree of squeezing. Noting that the degree
of squeezing in the squeezed vacuum input for $N=0.1$ is $46\%$, we see that
the squeezing in the resonance fluorescence output is substantially enhanced.

We plot $S_{X}$ against $\Omega $ and $\Delta $ in Fig. 3 where $%
N=0.125,\,\Phi =\pi ,$ and $\,\gamma =1$. This figure clearly shows that for
this value of $N,$ near maximal squeezing occurs over wide values of the
Rabi frequency $\Omega $ and detuning $\Delta $.

Figure 4 presents the total normally-ordered variance of the in-phase
quadrature of the fluorescent field, $S_{X}$, and the magnitude of the
atomic Bloch vector, $\Sigma =\langle \sigma _{x}\rangle ^{2}+\langle \sigma
_{y}\rangle ^{2}+\langle \sigma _{z}\rangle ^{2}$, for the parameters: $\Phi
=\pi ,\,\Delta =12.5\gamma ,\,\gamma =1$ and different squeezed photon
numbers (a) $N=0.05$, (b) $N=0.125$ and (c) $N=0.5$. The solid and dashed
lines represent $S_{X}$ and $\Sigma $ respectively. When $\Sigma =1$, the
atom is in a pure state. It is obvious that large squeezing in the resonance
fluorescence of the two-level atom occurs for pure atomic states %
\onlinecite{RFsq3,pures}. When $N=0.125$, maximal squeezing ($S_{X}=-0.25$)
is achieved at the Rabi frequency $\Omega =21.65\gamma $. The large
squeezing is due to the large atomic coherence in the pure state.

When eq. (\ref{cond3}) is satisfied, the atom is in the pure state (\ref{pur}%
) with $\alpha =0$. The corresponding total, normally-ordered variance $%
S_{X} $ of the in-phase quadrature of the fluorescent field is of the form

\begin{equation}
S_{X}^{PS}=\frac{N-M}{N+M+1/2}.  \label{sqpure}
\end{equation}
When $N=1/8,$ $M=3/8$. Then, from eq. (\ref{sqpure}) we have $%
S_{X}^{PS}=-0.25$ ($100\%$ squeezing). The corresponding value of the Rabi
frequency is $\Omega =\sqrt{3}\Delta $.

We plot $S_{X}^{PS}$, indicated by the solid line, against $N$ in Fig. 5,
which demonstrates that large squeezing occurs for small photon numbers. For
comparison, we also present the normally-ordered variance $S_{X}^{SV}$ of
the in-phase quadrature in the squeezed vacuum field, represented by the
dashed line in this figure. It is clear that the squeezing of the output
field (fluorescence) is greatly enhanced over the region $0<N\leq 0.562$,
compared with the squeezing of the input (squeezed vacuum) field. Hence, the
atom may be applied as a nonlinear optical element to amplify squeezing.

\subsection{$\frac{\protect\pi}{4}$-phase quadrature squeezing}

Fluorescent field squeezing can also occur in other phase quadratures with
the phase between $0$ (in-phase) and $\pi /2$ (out-of-phase)--- for example, 
$\theta =\pi /4$. The variance, $S_{\theta =\pi /4}$, of such a phase
quadrature is shown in Fig. 6, where $\gamma =1,\,\Phi =\pi /2,\,N=0.125$.
It is obvious that fluorescence squeezing, $S_{\pi /4}<0$, occurs over a
range of $\Omega ,\,\Delta $, and the optimal ($100\%$) degree of squeezing
takes place around $\Omega =0.612\gamma $ and $\Delta =0.25\gamma $.

As with in-phase quadrature squeezing, the optimal squeezing in the $\frac{%
\pi }{4}$-phase quadrature is also associated with a pure state (a highly
polarized atomic state). We plot $S_{\theta =\pi /4}$ and $\Sigma $ together
in Fig. 7 for the parameters: $\gamma =1,\,\Phi =\pi /2,\,\Delta =0.25$ and
different squeezed photon numbers (a): $N=0.05$, (b): $N=0.125$ and (c): $%
N=0.5$. This figure clearly shows again that a completely polarized atom can
emit a fluorescent field with large squeezing \onlinecite{RFsq3,pures}. When 
$N=0.125$, maximal squeezing ($S_{X}=-0.25$) is obtained at the Rabi
frequency $\Omega =0.612\gamma $.

We have shown that when $\Phi =\pi /2,\,\Delta =\Gamma -\gamma M$ and $%
\Omega =\gamma \sqrt{2M}/\left( \sqrt{N+1}+\sqrt{N}\right) $, the atom
develops into a stationary pure state, given by (\ref{pur}) with $\alpha
=\pi /4$. This highly polarized atom radiates a fluorescent field with $\pi
/4$-phase quadrature squeezing. The expression for the squeezing is same as
eq. (\ref{sqpure}), but for the $\pi /4$-phase quadrature. Perfect squeezing
is obtained for $N=0.125$, $\Delta =\gamma /4$ and $\Omega =\sqrt{6}\gamma
/4 $.

\subsection{Out-of-phase quadrature squeezing}

The resonance fluorescence exhibits out-of-phase (Y) quadrature squeezing
when $\Phi =0$ and $\Delta =0$ \onlinecite{RFsq2,RFsq3,RFsq4}. Furthermore,
if $\Omega $ obeys eq. (\ref{cond1}), the atom collapses into the pure state
(\ref{pur}) with $\alpha =\pi /2$. Consequently, the optimal Y-quadrature
squeezing, $S_{Y}^{PS}$, is given by eq. (\ref{sqpure}), as well. When the
squeezed photon number $N=0.125$ and the Rabi frequency $\Omega =\sqrt{3}%
\gamma /4$, the maximal degree of the Y-quadrature squeezing in the
resonance fluorescence is achieved.

\section{Summary}

We have shown that the total normally-ordered variance of the phase
quadrature of the fluorescent field emitted from a coherently driven
two-level atom interacting with a squeezed vacuum can be greatly (or
completely) squeezed. The squeezing in the fluorescent field is greatly
increased for small values of $N$ compared with the degree of squeezing of
the input squeezed vacuum field. Therefore, squeezing may be amplified
through resonance fluorescence. The squeezing in the output may indeed be
the maximum obtainable, when $N=1/8$. Analytic expressions for the
conditions for maximum squeezing are obtained in this case. Depending upon
the values of the squeezed phase $\Phi $ and the detuning $\Delta $,
squeezing can occur in the in-phase, out-of-phase, or any other phase
quadrature of the resonance fluorescence. Large squeezing is always
associated with the atom evolving into a pure state (a highly polarized
atomic state).

From the experimental point of view, a cavity configuration may be the best
candidate to investigate the atom/squeezed-vacuum interactions \cite{sqcav}.
As shown in Refs. \onlinecite{sqcav,badc1}, many squeezing-induced effects
in free space can carry over to the cavity situations in the bad cavity
limit, where the atom evolves in accordance with formally the same equations
as those in free space. We expect that large squeezing in resonance
fluorescence will still take place in the cavity configuration.

\acknowledgments
This work is supported by the United Kingdom EPSRC. We are grateful to S.
Bali for sending us a preprint of his work.

\begin{figure}[tbp]
\caption{The polar angles $\protect\alpha$ and $\protect\beta$ defining the
position of the Bloch vector ${\bf B}$ on the Bloch sphere.}
\label{fig1}
\end{figure}

\begin{figure}[tbp]
\caption{The total normally-ordered variance of the in-phase quadrature, $%
S_{X}$, against the Rabi frequency $\Omega $ and the squeezed phase $\Phi /%
\protect\pi $, for $\protect\gamma =1,\,\Delta =10,\,N=0.1$.}
\label{fig2}
\end{figure}

\begin{figure}[tbp]
\caption{Same as FIG. 2, but against $\Omega$ and $\Delta$, with $\protect%
\gamma=1,\,\Phi=\protect\pi,\, N=0.125$.}
\label{fig3}
\end{figure}

\begin{figure}[tbp]
\caption{$S_{X}$ and $\Sigma$ as functions of $\Omega$, for $\protect\gamma %
=1,\,\Phi=\protect\pi,\,\Delta=12.5$ and (a): $N=0.05$, (b): $N=0.125$ and
(c): $N=0.5$. The solid and dashed lines represent respectively $S_{X}$ and $%
\Sigma$.}
\label{fig4}
\end{figure}

\begin{figure}[tbp]
\caption{$S_{X}^{PS}$ and $S_{X}^{SV}$ as functions of $N$, represented by
the solid and dashed lines respectively. }
\label{fig5}
\end{figure}

\begin{figure}[tbp]
\caption{Same as FIG. 2, but the variance of the $\protect\pi/4$-phase
quadrature, $S_{\protect\theta=\protect\pi/4}$, with $\protect\gamma%
=1,\,\Phi=\protect\pi/2,\, N=0.125$.}
\label{fig6}
\end{figure}

\begin{figure}[tbp]
\caption{$S_{\protect\theta=\protect\pi/4}$ and $\Sigma$ as functions of $%
\Omega$, for $\protect\gamma =1,\,\Phi=\protect\pi/2,\,\Delta=0.25$ and (a):$%
N=0.05$, (b): $N=0.125$ and (c): $N=0.5$. The solid and dashed lines
represent respectively $S_{\protect\pi/4}$ and $\Sigma$.}
\label{fig7}
\end{figure}

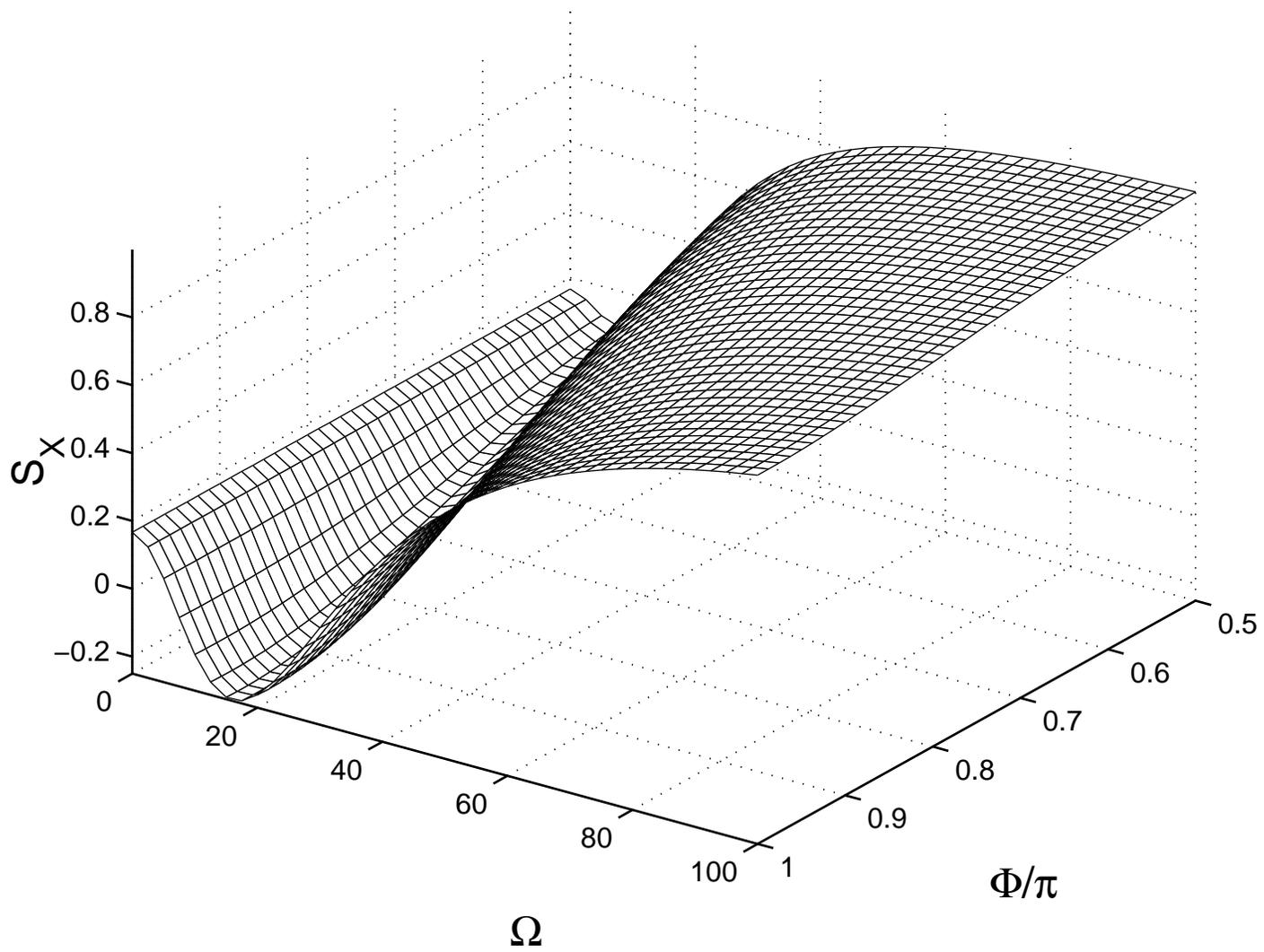
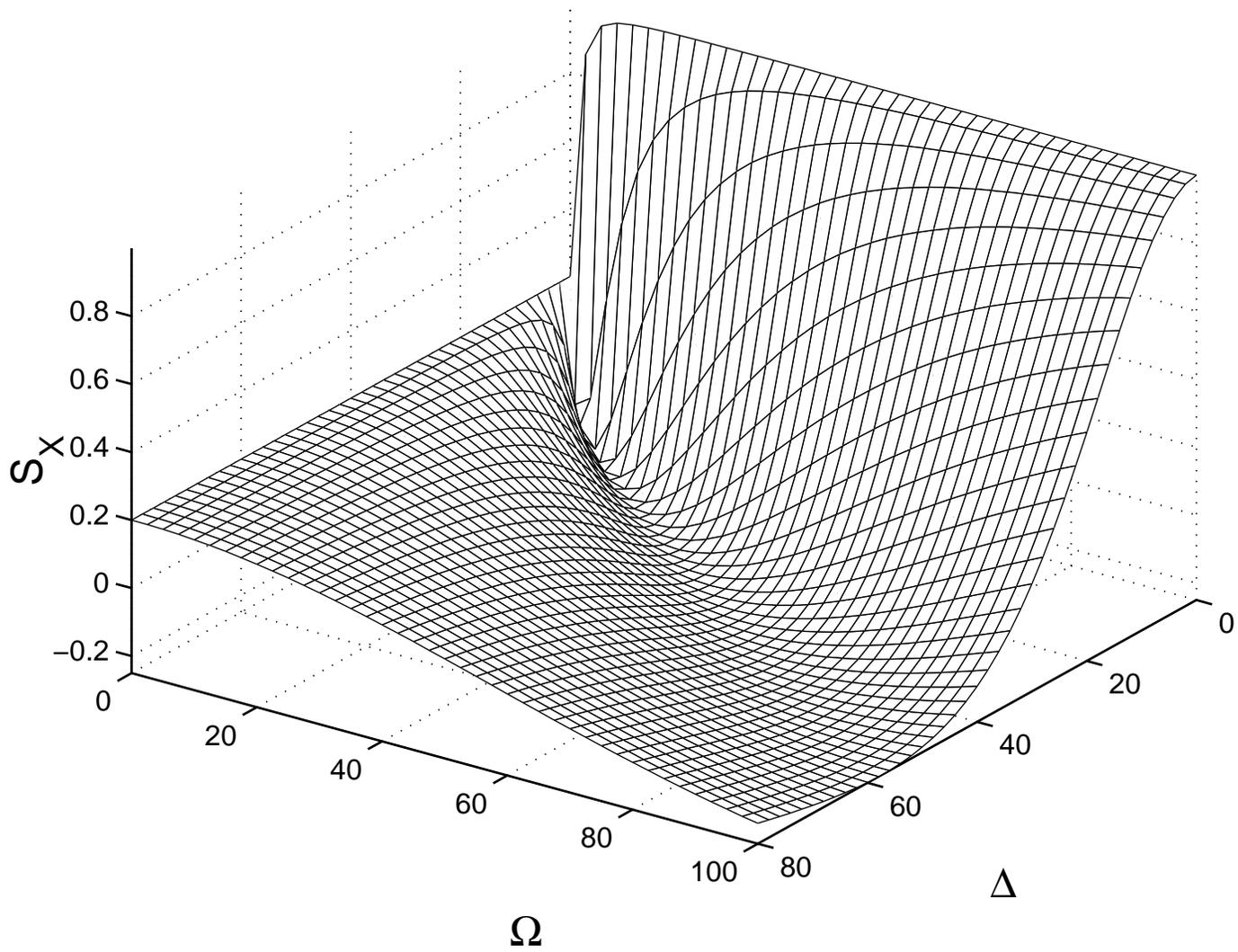
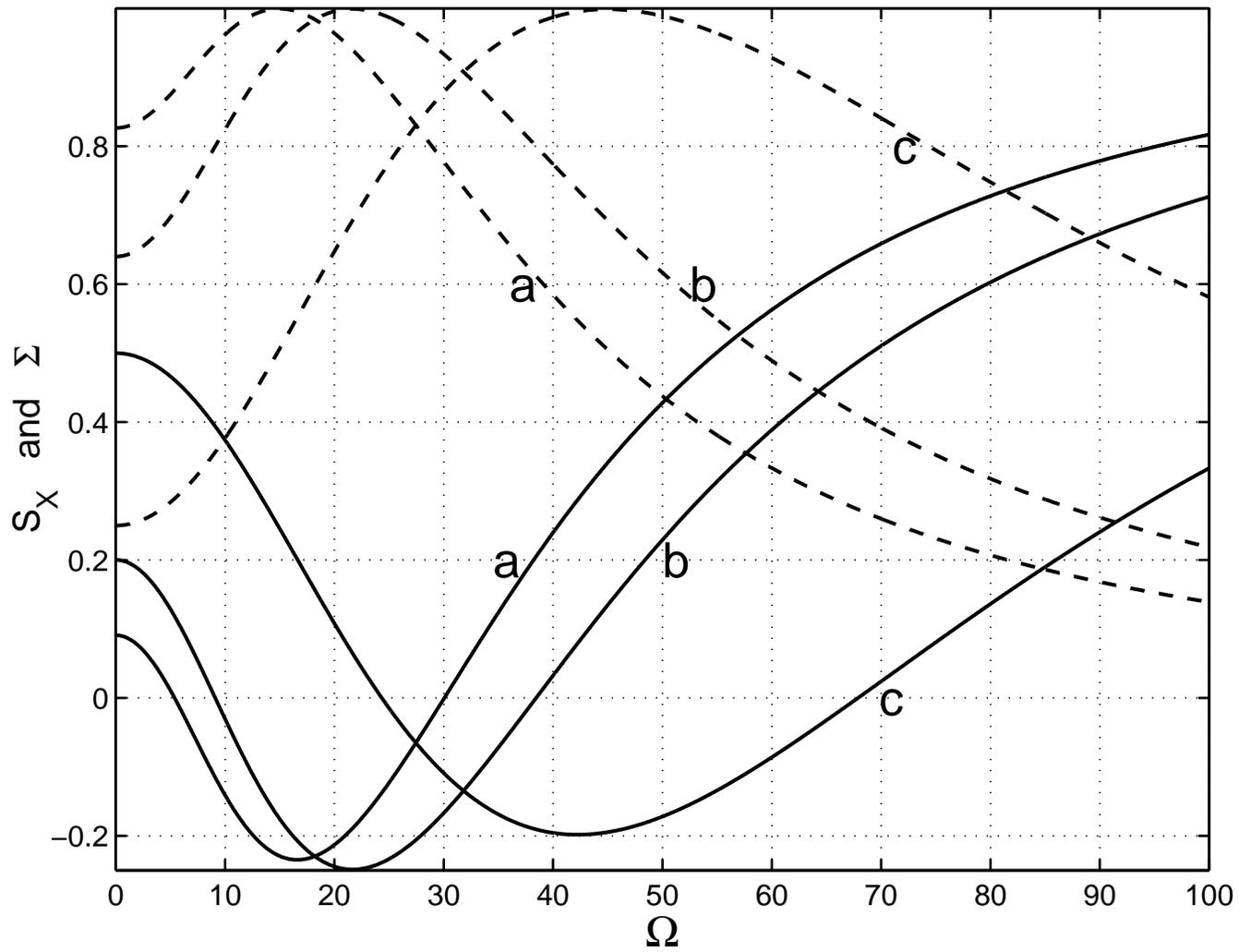
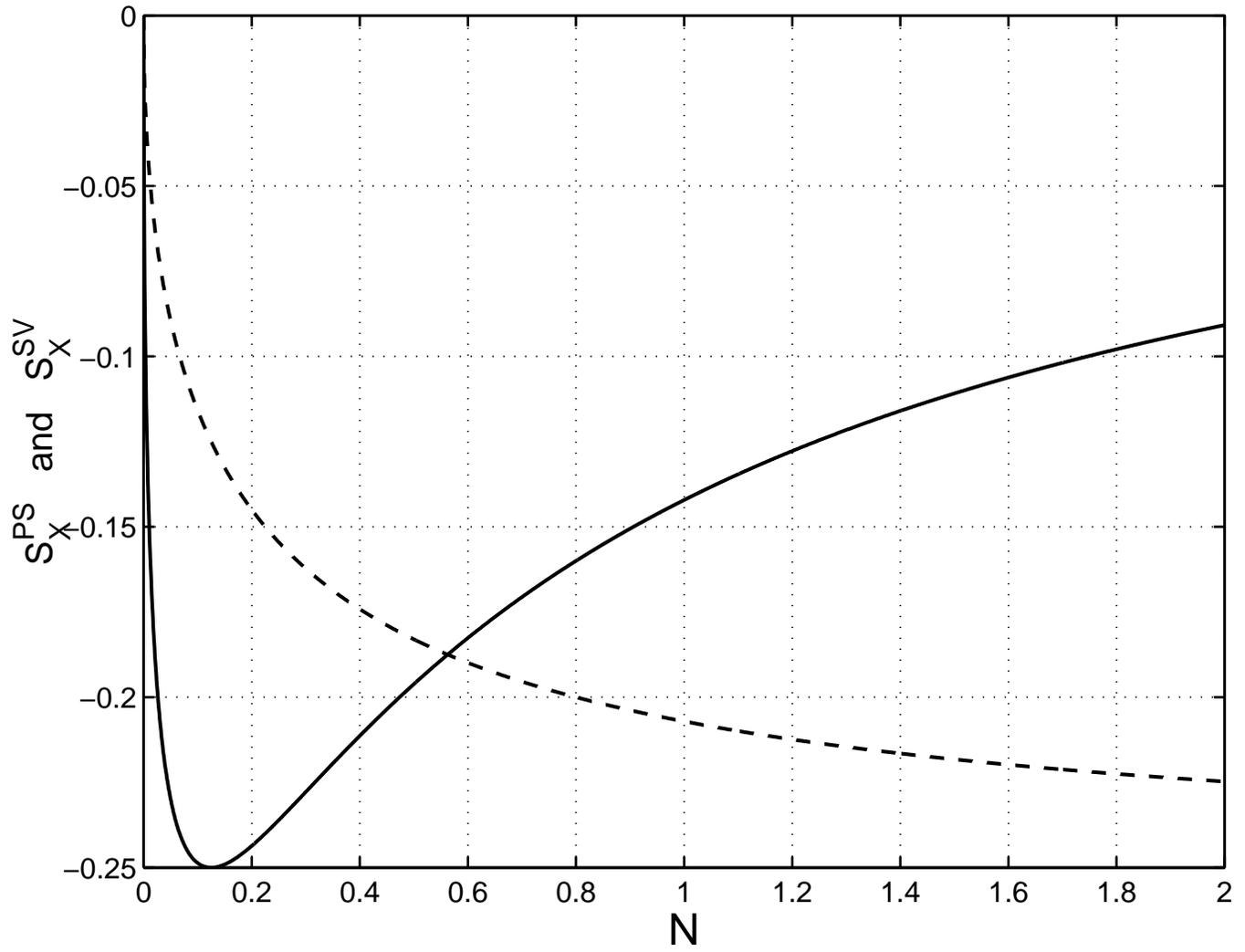
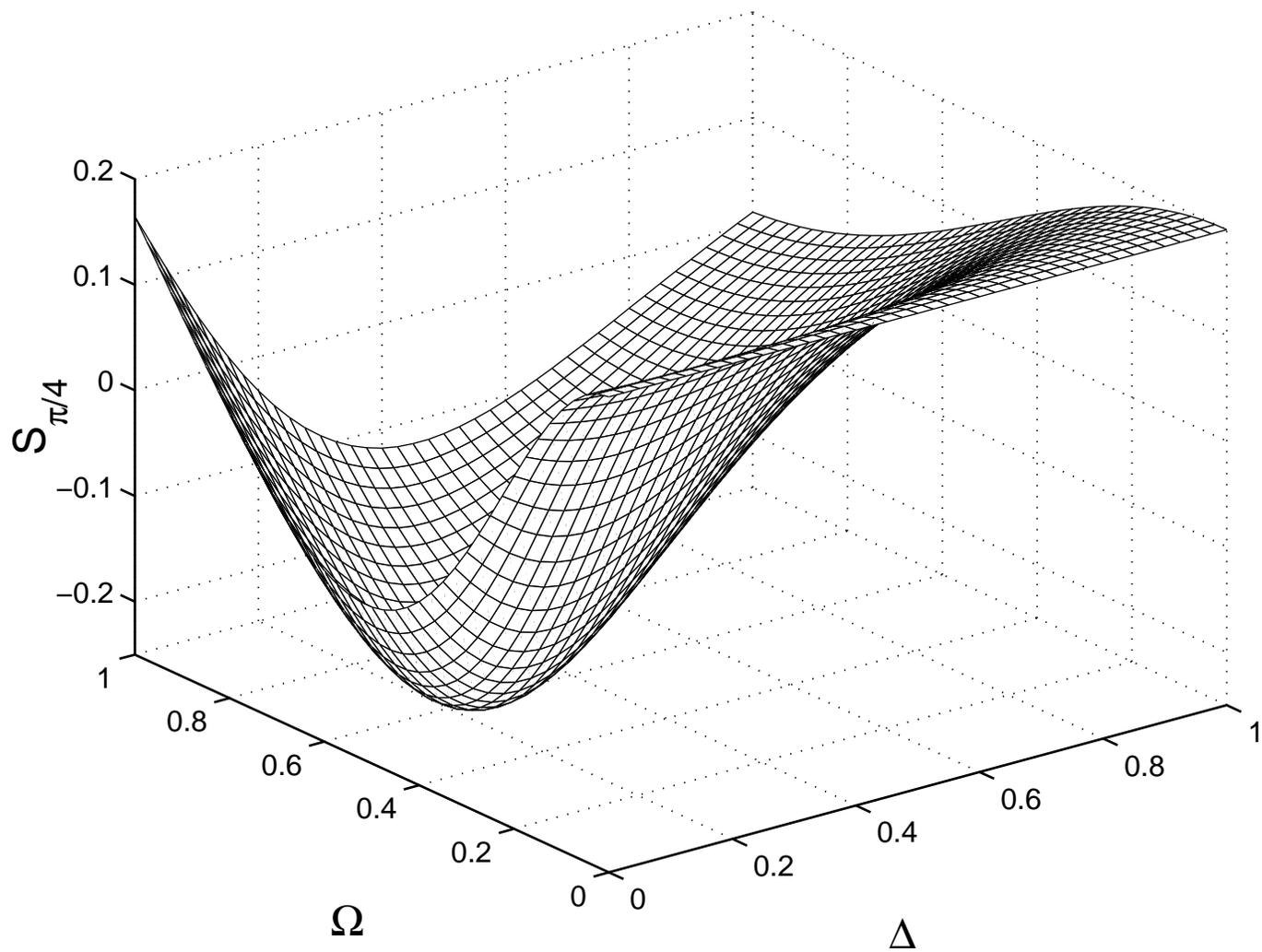
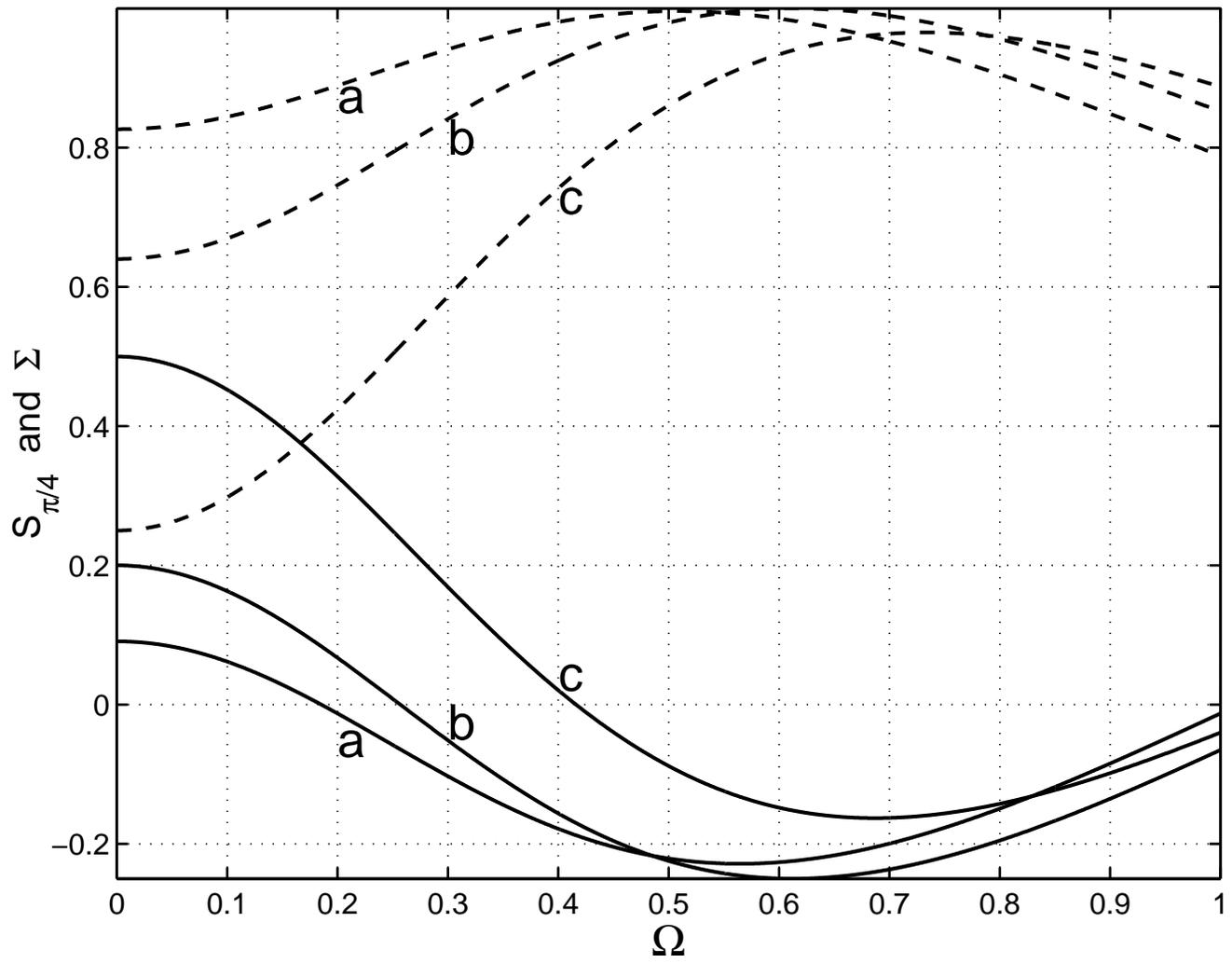
\end{document}